\documentclass[article]{JHEP3}

\usepackage{amsmath}
\usepackage{amssymb}
\usepackage{amsfonts}
\usepackage{mathtools}
\usepackage{cite}
\usepackage{graphicx}
\usepackage{amsthm}
\usepackage{booktabs}
\usepackage{multirow}
\usepackage{mathrsfs}
\usepackage{dsfont}
\usepackage{gensymb}
\usepackage{comment}
\allowdisplaybreaks

\def\beq{\begin{equation}}

\def\eeq{\end{equation}}

\relax

\title{A new regime of the C-metric: Super accelerating black holes}
\author{Francisco Silva$^a$
\\
$^{a}$
Departamento de F\'\i sica,\\ 
Faculdade de Ci\^encias da Universidade de Lisboa,\\
Campo Grande, 1749-016 Lisboa, Portugal\\
\email{fc58297@alunos.ciencias.ulisboa.pt}
}

\abstract{We define and investigate the super accelerating regime of the charged C-metric, which arises at large acceleration. In this regime, the polar coordinate must be restricted to a smaller interval within $(0,\pi)$, and the resulting boundary defines a new axis of symmetry. This changes the conical singularity structure of the spacetime and, consequently, its string/strut interpretation. We further show that, when the conical defect at the south pole is removed, the Komar mass takes the simple form $M_K=1/(4A)$, up to the normalization of the time translation Killing vector. Finally, we explore the near extremal limits of this regime, identifying new limiting geometries.}

\begin{document}

\maketitle

\vfill

\eject

\section{Introduction}
\noindent
The C-metric has been the subject of many studies. It was first found by Levi-Civita \cite{Levi-Civita:1917}, independently discovered by Weyl \cite{Weyl:1917gp} and later interpreted by Kinnersley and Walker in \cite{Kinnersley:1970zw} as describing uniformly accelerating black holes. 
Subsequently, the geometrical and physical properties of the spacetime, including its causal structure, conformal diagrams, horizon structure, and the interpretation of the conical singularities responsible for the acceleration, were studied in detail by Bonnor \cite{Bonnor:1983}, Hong and Teo \cite{Hong:2003gx, Hong:2004dm}, and Griffiths and Podolský \cite{Griffiths:2006tk}, while its generalizations to including a cosmological constant and extremal limits were performed by Dias and Lemos \cite{Dias:2002mi, Dias:2003xp, Dias:2003up}. More recently, a complete analysis of the global structure of the C-metric and its extensions has been presented in \cite{Cembranos:2022dhi}.

Lately, accelerating black holes have attracted renewed interest due to their thermodynamic properties which have been studied in detail in several works \cite{Hale:2025veb, Appels:2016uha, Abbasvandi:2018vsh, Luo:2024cwm}.

Additionally, the propagation of fields on accelerating black hole backgrounds has become another active area. Its lensing properties have been analyzed in \cite{Frost:2020zcy}, while studies of field perturbations permitted the study of quasinormal modes, superradiance and the stability of its Cauchy horizon  \cite{Destounis:2020pjk, Lei:2023mqx, Destounis:2022rpk, Moura:2026ken, BarraganAmado:2023wxt, Chen:2024rov}.

In most of these works however, a constraint on the acceleration parameter $A$ has been imposed. The argument is that allowing for a large parameter $A$ can introduce pathologies on the angular part of the spacetime due to a signature change on the metric. Not only that, but the horizon associated with the acceleration of the black hole can swap places with the previous event horizon of the black hole. This is precisely how we will define the super accelerating regime and has already been identified in \cite{Frost:2020zcy} and \cite{Cembranos:2022dhi}, 
but deemed unphysical by the first. 
On the other hand, the work of \cite{Cembranos:2022dhi} has correctly identified and studied the global structure of the uncharged super accelerating black hole, as well as identified the necessary restriction on the polar coordinate range. However, many other characteristics of the super accelerating regime have been left unexplored.

On this work, we will consider the super accelerating regime of the charged C-metric and explain how the problems mentioned above can be easily dealt with. The inclusion of charge divides the super accelerating regime into two parts depending on how large the acceleration parameter is. The article is organized as follows. In section \ref{A brief review of the C-metric} we give a brief review of the charged C-metric, explain why the parameter $A$ is interpreted as the acceleration of the black hole and what sources this acceleration. In section \ref{Horizon structure} we define the different accelerating regimes. Further, in each case we study the properties of the new axis of symmetry, the conical singularities, conformal infinity and the topology of the spacetime. Later in \ref{Super accelerating regime with cosmological constant} we study how one can construct naked singularities by adding a negative cosmological constant to the spacetime. From then on, we study the geometrical and physical properties of the super accelerating regime in section \ref{Geometrical and physical quantities}. There, we compute the Komar mass, charge and area of the black hole, where we find that in the super accelerating regime, the Komar mass is simply given by $M_K=1/(4A)$, provided that the conical singularity on the south pole is removed. Finally, in section \ref{Near extremal}, we review the known extremal limits of the C-metric and obtain the new super accelerating extremal regimes. To end, we conclude in section \ref{Conclusion} with the main results obtained as well as possible questions to explore in the future in regard to super accelerating black holes.

\section{A brief review of the C-metric} \label{A brief review of the C-metric}
\noindent
The four-parameter charged C-metric with cosmological constant provides an exact solution of the Einstein--Maxwell equations describing an accelerating black hole. We parameterize the solution by the mass and charge parameters
$M$ and $Q$, the acceleration $A$, and the cosmological constant $\Lambda$.
The distinction between the parameters $M,Q$ and the corresponding physical mass and charge will be discussed in Sec.~\ref{Geometrical and physical
quantities}. We allow $\Lambda$ to be positive, zero, or negative, and for $\Lambda<0$ we introduce the AdS length scale $L=\sqrt{-3/\Lambda}$.

In the Hong-Teo coordinates \cite{Hong:2003gx}, the metric takes the form
\begin{equation}
    ds^2=\frac{1}{\Omega^2(r,\theta)}\left(-f(r)dt^2+\frac{dr^2}{f(r)}+\frac{r^2d\theta^2}{P(\theta)}+P(\theta)r^2\sin^2{\theta}d\varphi^2\right),
    \label{Metric}
\end{equation}
where metric functions are defined as
\begin{equation}
\begin{aligned}
    f(r)&=\left(1-\frac{2M}{r}+\frac{Q^2}{r^2}\right)\left(1- A^2r^2\right)-\frac{\Lambda}{3}r^2, \\ P(\theta)&= 1-2AM\cos{\theta}+A^2Q^2\cos^2{\theta},\\\Omega(r,\theta)&=1-Ar \cos{\theta},
\label{metric functions}
\end{aligned}
\end{equation}
and their input on the properties of the black hole will be discussed in detail in section \ref{Horizon structure}.

The electric four-potential $\mathscr{A}$  and its associated field strength $F=d\mathscr{A}$, that realize this black hole solution are given by
\begin{equation}
    \mathscr{A}=-\frac{Q}{r}dt, \quad F=\frac{Q}{r^2}dr\wedge dt.
\end{equation}

\subsection{The interpretation and source of acceleration} \label{Interpretation and source of acceleration}
\noindent
In this subsection, we aim to motivate the physical interpretation of the parameter $A$ as acceleration and discuss what sources it.

To address this first problem, we simplify the black hole metric by taking $M=Q=\Lambda=0$. Here, the metric becomes
\begin{equation}
    ds^2=\frac{1}{(1-Ar\cos\theta)^2}\left(-(1-A^2r^2)dt^2+\frac{dr^2}{1-A^2r^2}+r^2d\theta^2+r^2\sin^2\theta d\varphi^2 \right).
\end{equation}
Consider an observer described by timelike world lines of constant spatial position. We shall look for what is the acceleration that such observer experience as a function of the parameter $A$ and position $r$.

Its four velocity is given by $u_\mu=(u_t,0,0,0)$, where normalization imposes $u_\mu u^\mu=-1$. To compute the acceleration four vector $a_\mu$, we use the covariant derivative $\nabla_\nu$ and Christoffel symbol $\Gamma^t_{t \mu}$, such that
\begin{equation}
    a_\mu=(u^\nu\nabla_\nu)u_\mu=-u_tu^t \Gamma_{t\mu}^t=\frac{1}{2}\partial_\mu \log(-g_{tt})=\left(0,-\frac{A^2r}{1-A^2r^2}+\frac{A\cos{\theta}}{1-Ar\cos\theta},-\frac{Ar\sin{\theta}}{1-Ar\cos{\theta}},0 \right).
\end{equation}
Therefore, we conclude that the norm of the acceleration experienced by such observer given by
\begin{equation}
    a=\sqrt{a^\mu a_\mu}=\frac{A\left(1-Ar\cos{\theta}\right)}{\sqrt{1-A^2r^2}}.
\end{equation}
and that it is orthogonal to its four velocity, since $u^\mu a_\mu=0$.
Consequently, we deduce that the acceleration of a static observer at the origin of the spacetime ($r=0$) is $a=A$. Thus we have interpreted the parameter $A$ as acceleration.

Regarding the second problem about the source of acceleration, we should take back the parameters $M,Q,\Lambda,A\neq0$ and look at the conical singularities of the spacetime. These singularities can be identified at the north ($\theta=0$) and south ($\theta=\pi$) poles and represent points where an infinitesimal small circle has $\text{Circumference}$ different from $2\pi\times\text{Radius}$. To analyze their presence on the C-metric, we look for the ratio of these two quantities on both poles. While doing so, we keep an open mind regarding the period of the variable $\varphi$, by introducing the parameter $C$, such that $\varphi\in (0,2\pi C)$. We find that, on $\theta=0$
\begin{equation}
    \lim_{\theta \rightarrow 0} \frac{\text{Circumference}}{2\pi \text{ Radius}}=\lim_{\theta \rightarrow 0}\frac{\int_0^{2\pi C}r\sqrt{P(\theta)}\sin\theta d\varphi}{2\pi\times \int _0^\theta r/\sqrt{P(\vartheta)}d\vartheta}=\lim_{\theta \rightarrow 0}\frac{CP(\theta) \sin\theta}{\theta}=CP(0),
    \label{stringCP(0)}
\end{equation}
while on $\theta=\pi$
\begin{equation}
        \lim_{\theta \rightarrow \pi} \frac{\text{Circumference}}{2\pi \text{ Radius}}=\frac{CP(\theta)\sin\theta}{\pi-\theta}=CP(\pi).
    \label{conical singularity}
\end{equation}

This tells us that in general there is a conical singularity at each pole of the black hole. However, one of such can be removed, by choosing an appropriate value for $C$.

It follows that if one chooses to remove the conical singularity at the south pole ($\theta=\pi$), we impose
\begin{equation}
    C=\frac{1}{P(\pi)}=\frac{1}{1+2MA+A^2Q^2}\eqcolon C_\pi.
    \label{Cpi}
\end{equation}
However, this introduces a deficit angle at $\theta=0$ given by
\begin{equation}
    \delta_0^\pi=2\pi\left(1-\frac{P(0)}{P(\pi)}\right)=8\pi\frac{AM}{1+2MA+A^2Q^2}=8\pi AMC_\pi.
    \label{delta0}
\end{equation}
\footnote{
Clarifying the notation for the angle defect: The upper script for $\delta$ denotes the conical singularity we have removed, while the subscript indicates the angle defect we are evaluating.}Consequently, this configuration is interpreted as a cosmic string of tension $\mu_0=\delta_0^\pi/8\pi$ that is pushing the black hole and hence sourcing its acceleration.

On the other hand, we could have identically removed the conical singularity at the north pole ($\theta=0$) by choosing 
\begin{equation}
    C=\frac{1}{P(0)}=\frac{1}{1-2MA+A^2Q^2}\eqcolon C_0,
    \label{C0}
\end{equation}
but this would create an angle defect at $\theta=\pi$ with 
\begin{equation}
    \delta_\pi^0=2\pi\left(1-\frac{P(\pi)}{P(0)}\right)=-8\pi\frac{AM}{1-2MA+A^2Q^2}=-8\pi AM C_0.
    \label{deltapi}
\end{equation}
The sign of the angular defect $\delta_\pi^0$ appears to depend on the black hole parameters. However, as we will see in \ref{Horizon structure}, configurations with $\delta_\pi^0>0$ should not be considered. Thus, restricting to $\delta_\pi^0<0$, we find a cosmic strut pushing the black hole, with tension $\mu_\pi=\delta_\pi^0/8\pi$.
Comparing this with the cosmic string configuration along $\theta=0$, we see that cosmic strings have positive energy density and exert negative pressure, whereas cosmic struts have negative energy density and exert positive pressure. In both cases, it is the string or strut that provides the force responsible for accelerating the black hole towards the north. Finally, reversing the direction of acceleration, $A\to-A$, is equivalent to reversing the orientation of the black hole, $\theta\to\pi-\theta$, since the combination of these two transformations is an isometry of the C-metric \eqref{Metric}.


\section{Horizon structure\label{Horizon structure}}
\noindent
We now proceed to interpret the properties of the spacetime geometry dictated by the functions $\Omega(r,\theta)$, $f(r)$ and $P(\theta)$. The cosmological constant will be first turned off ($\Lambda=0$), but then turned back on again on section \ref{Super accelerating regime with cosmological constant}. 

Let us start by understanding the horizons of the black hole. These are given by the zeros of the function $f(r)$,
\begin{equation}
    r_\pm=M\pm \sqrt{M^2-Q^2},\quad r_{\pm A}=\pm\frac{1}{A}.
\end{equation}

We notice that the horizons of the charged black hole solution, also known as the Reissner Nordstr\"{o}m black hole, are preserved when acceleration is included in the spacetime\footnote{This is not true for a non zero cosmological constant ($\Lambda\neq0$).}, so we naturally impose $M>|Q|$. Additionally, there is a new physical horizon $r_A\coloneqq r_{+A}=1/A$ and a virtual horizon $r_{-A}=-1/A$ which is deemed unphysical.

As one can see so far, the contribution of the black hole parameters to its horizons are split. This means that the parameter $A$ does not contribute to $r_\pm$ and the parameters $M$ and $Q$ do not contribute to $r_A$. This leads us to impose constraints on their values by hand.

What is usually considered, is to say that, physically speaking, the acceleration of the black hole should be small such that
\begin{equation}
    r_+<r_A.
\end{equation}
However, \textit{a priori}, even though it might seem unphysical, it is theoretically possible to conceive a configuration where 
\begin{equation}
    r_-<r_A<r_+ \quad \text{or even}\quad r_A<r_-<r_+.
\end{equation}
To the first configuration we call the first super accelerating regime, while for the second configuration we call the second super accelerating regime. It is worth noting that in the region between the outer horizon and the cosmological horizon\footnote{We opt to classify the horizons in an ascending order, as the inner horizon, the outer horizon and the cosmological horizon. Alternatively we will also call it the Cauchy horizon and event horizon respectively for the first two.}m the coordinate $r$ remains timelike, since $f(r)>0$ there.

The purpose of the following sections will be to discuss the physical properties of these different parameter configurations and how they change our interpretation of the C-metric. The discussion will be split into four parts, depending the values of the black hole parameters. For the first three subsections \ref{Common accelerating regime}, \ref{First super accelerating regime} and \ref{Second super accelerating regime}, $\Lambda$ will be set to zero, but then turned back on again in \ref{Super accelerating regime with cosmological constant}. The focus will be on the additional features the angular function $P(\theta)$ introduces, how the string/strut picture changes, the new properties the conformal factor $\Omega(r,\theta)$ creates and what is the topology of the spacetime.

Before doing so, it is convenient to introduce two features. Let us define the surface gravity of the black hole as
\begin{equation}
    k_\pm =\frac{f'(r_\pm)}{2},\quad k_{\pm A}=\frac{f'(r_{\pm A})}{2}.
\end{equation}
A curious property regarding the periodicity of the angular variable $\varphi$, fixed by the conical singularity we choose to remove, is noticed by the fact that we are allowed to rewrite \eqref{Cpi} and \eqref{C0} as 
\begin{equation}
    C_\pi=\frac{A}{k_{-A}},\quad C_0=\frac{-A}{k_A}.
    \label{Csurfacegravity}
\end{equation}

Even though the value of $C$ originally comes from the angular function $P(\theta)$ evaluated at the poles of the black hole, one notices a connection with the function $f(r)$ as well. In particular, it is quite intriguing to see that the string configuration corresponding to choosing $C=C_\pi$, is associated with the surface gravity of the virtual horizon $r_{-A}$. However, one might also deem this connection as just a coincidence, since considering $\Lambda\neq0$ changes the story completely.

The other property we wish to examine before analyzing each accelerating regime separately is that of conformal infinity. This is a spacetime region where the conformal factor vanishes, $\Omega(r,\theta)=0$. As one can notice by \eqref{Metric}, the metric becomes divergent in all of its components while its inverse reduces to zero.
Further, if we consider $Q=0$ for simplicity, we find that the only identically non zero component of the curvature tensor actually vanishes on conformal infinity \cite{Griffiths:2006tk}. Therefore, we are led to interpret this boundary as flat spacetime region where physical points become infinitely distant from one another and all spatial directions become isotropic. Additionally this region separates the spacetime into two, imposing that we must focus on either $\Omega(r,\theta)>0$ or $\Omega(r,\theta)<0$ regions. %

\subsection{Regular accelerating regime ($r_-<r_+<r_A$)} \label{Common accelerating regime}
\noindent
The regular accelerating regime is naturally defined as the configuration where $r_-<r_+<r_A$. Therefore, $r_-$ is the Cauchy horizon, $r_+$ the event horizon and $r_A$ the cosmological horizon.

On this regime, the function $P(\theta)$ mainly contributes to the conical defect story. Let us then, proceed to analyze the string/strut configurations of this black hole. Consider the cosmic string configuration associated with the removal of the conical singularity at $\theta=\pi$. A natural condition a sensible spacetime must have is that $\delta_0^\pi<2\pi$, since $\delta_0^\pi=2\pi$ would mean the the entire azimuthal angular direction would collapse. As such, this condition imposes constraints on the the black hole parameters, since
\begin{equation}
    \delta_0^\pi<2\pi \iff 4MA<1+2MA+A^2Q^2\iff r_A^2-2Mr_A+Q^2>0\iff r_A>r_+ \vee r_A<r_-. 
    \label{delta0 condition}
\end{equation}
It would seem that this rules out the first super accelerating regime, however there will be a natural way out discussed in \ref{First super accelerating regime}. On the other hand, we check that the regular and second accelerating regimes are free from those pathologies.

Looking at the strut configuration, associated with the removal of the conical singularity at $\theta=0$, we notice that $\delta_\pi^0$ in \eqref{deltapi} must now be negative. There are two quick and elegant ways to see this. One is by looking at the denominator of $C_0$ and noticing the following set of equalities:
\begin{equation}
    1-2MA+A^2Q^2=A^2(r_A^2-2Mr_A+Q^2)=A^2(r_A-r_+)(r_A-r_-),
    \label{Cpi simplification}
\end{equation}
which should be greater than zero in this regime. The other is to look into \eqref{Csurfacegravity} and notice that, since $r_A$ is now the cosmological horizon, then $k_A<0$ and therefore $C_0>0$. We also notice that in each accelerating regime, the sign of $\delta_\pi^0$ changes\footnote{This would correspond as having a cosmic string on $\theta=\pi$ pulling the black hole in the first super accelerating regime, but this would be pathological, since the string would be pushing the black hole in the exact opposite way we had found it was moving before.}. However in \ref{First super accelerating regime} we will explain why on that regime the conical singularity at $\theta=0$ should not be removed and hence it will be eliminated at $\theta=\pi$.

Additionally, for the strut configuration, one does not necessarily get a geometric constraint $-2\pi<\delta_\pi$ to avoid singular behaviors of the angular coordinates. Therefore, this configuration does not impose new constraints on the black hole parameters. Further, struts are usually deemed more unphysical than strings, due to their negative energy density, therefore, the string configuration is more commonly chosen.

To end this section, we will discuss when conformal infinity is reached. This is satisfied by the coordinate pair $(r,\theta)$ such that
\begin{equation}
    1-Ar\cos\theta=0\iff r\cos\theta=r_A.
\end{equation}
Since in this regime, a timelike coordinate $r$ should be enclosed by the event and cosmological horizons of the black hole $r\in (r_+,r_A)$, and due to $\cos{\theta}\in(-1,1)$, we note that conformal infinity is only reached on the single spatial point given by $(r,\theta)=(r_A,0)$. 

Finally, the topology of this regime is essentially conformal to 2-dimensional Minkowski times a deformed 2-sphere $\tilde S^2$, $\mathbb{M}_2\times \tilde S^2$, except for a spacetime point which corresponds to a string or strut extending to infinity. Additionally, there is another exception which corresponds to the topology of the acceleration horizon $r_A$. In \cite{Luo:2024cwm} it is argued that the topology of the acceleration horizon is $\mathbb{R}_t\times \mathbb{R}^2=\mathbb{R}^3$ instead of $\mathbb{R}_t\times \tilde S^2$ and this happens due to the non compactness introduced reaching conformal conformal infinity.

To end this section, we leave figure \ref{Figure common accelerating regime} which summarizes the horizon structure we have just discussed. However, some care is required, as we are representing constant $r$ surfaces as spheres while the black hole is not really spherically symmetric.

\FIGURE{
    \centering
    \includegraphics[width=0.6\linewidth]{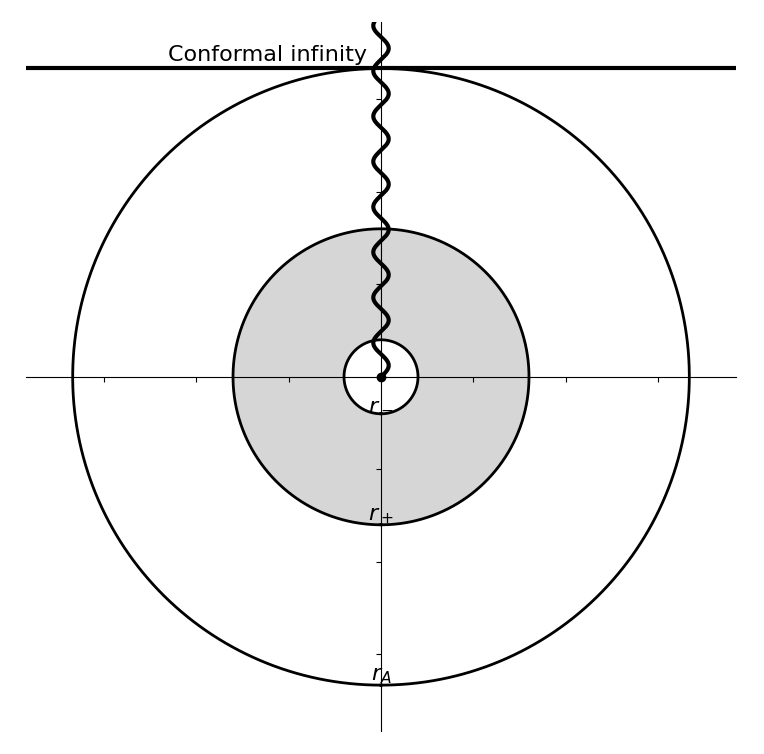}
    \caption{Two dimensional representation of the C-metric in the regular accelerating regime. On the vertical axis we have the $z=r \cos\theta$ and on the horizon axis we have $x=r\sin\theta$. The horizons of the black hole are plotted and the shaded regions represent when the coordinates $t,r$ change signature. $r_-$ is the Cauchy horizon, $r_+$ the event horizon and $r_A$ the cosmological horizon. Conformal infinity is represented by the constant $z=r_A$ line and the conical defect is chosen to lay on the $\theta=0$ line representing a cosmic string.}
    \label{Figure common accelerating regime}}



\subsection{First super accelerating regime ($r_-<r_A<r_+$)} \label{First super accelerating regime}
\noindent
The first super accelerating regime is achieved when the horizons of the black hole are ordered as $r_-<r_A<r_+$. Meaning that $r_-$ is the Cauchy horizon, $r_A$ the event horizon and $r_+$ the cosmological horizon of the black hole.

Let us start by analyzing the function $P(\theta)$. The reason this accelerating regime is usually disregarded on the literature is because there are now some $\theta$, such that the function $P(\theta)<0$. This is usually deemed pathological, since it alters the metric signature \eqref{Metric}. We will further describe what kind of pathologies these are, but for now, let us focus on how to remedy this. We define $\theta_+=\arccos \left(\frac{r_A}{r_+}\right)$ as the zero of $P(\theta)$ (i.e. $P(\theta_+)=0$) and so $P(\theta)<0$ for $\theta\in (0,\theta_+)$. A natural and easy way to trespass this difficulty is by restricting $\theta \in \left(\theta_+,\pi\right)$. At first sight, this suggests that there must be an analogous of an horizon on the angular coordinate, which we could tentatively call angular horizon, set by $\theta=\theta_+$, but its nature shall be analyzed in more detail.

To understand this boundary more clearly, note that at $\theta=\theta_+$, the metric component $g_{\theta \theta}$ diverges while the metric component $g_{\varphi \varphi}$ vanishes. This gets the interpretation that while local polar distances get infinitely long there, angular $\varphi$ distances shrink to zero, essentially signaling that all the azimuthal angular direction collapse onto a single spacetime point. To be more precise on this last statement, notice that a constant $\theta$ and closed $\varphi$ curve has length 
\begin{equation}
    \ell_{\varphi}(\theta)=\frac{2\pi C r\sin\theta\sqrt{P(\theta)}}{\Omega(r,\theta)}.\label{length varphi}
\end{equation} 
We therefore find that $\ell_{\varphi}(\theta) \to 0$ as $\theta\to \theta_+$. This is analogous to the sphere geometry where the length of a constant $\theta$ curve goes to zero as one of the poles is reached. Additionally, we investigate the length of a constant $\varphi$ curve starting at some $\theta_0$ and ending at $\theta_+$. We find that the length of this curve is given by
\begin{equation}
\ell_{\theta}=\int_{\theta_0}^{\theta_+}\frac{r}{\Omega(r,\theta)\sqrt{P(\theta)}}d\theta.\label{length theta}
\end{equation}
For a generic non vanishing conformal factor\footnote{The case where the conformal factor is zero is analyzed at the end of this subsection.}, and considering that generically $\theta_+$ is a simple zero of $P(\theta)$ (i.e. $P(\theta) \propto (\theta-\theta_+)$, when $\theta\sim \theta_+$) for non extremal configurations\footnote{The zeros of $P(\theta)$ become degenerate when $r_+=r_-$.}, we surprisingly find that while locally $g_{\theta \theta}\to \infty$, $\ell_{\theta}< \infty$ converges. We find that in fact the angular part of the metric forms a compact surface, conformal to the topologically equivalent of a deformed 2-sphere, since $\ell_\varphi (\theta_+)=\ell_{\varphi}(\pi)=0$ and $\ell_\varphi<\infty$. Moreover, the region set by $\theta=\theta_+$ can be interpreted as an axis of symmetry of this 2-sphere. This is easily seen by noticing that the azimuthal Killing vector $(\partial_\varphi)$ becomes null there. The topology of this regime will introduce important consequences on the mass of the C-metric analyzed in section \ref{Geometrical and physical quantities} and the conical singularities of the spacetime which shall be discussed on this subsection. 


For now let us analyze whether the interpretation of the $\theta=\theta_+$ as a new kind of horizon would lead somewhere. We note that while the metric becomes degenerate in this region, the curvature invariants remain finite, which signals that this divergences is just but a coordinate artifact. Also, we might ask whether geodesics ever trespass the $\theta=\theta_+$ boundary. 

To do so, we start with the most general motion of a null geodesic. The geodesic equations are separable and yield \cite{Frost:2020zcy}
\begin{equation}
    \frac{dt}{d\eta}=\frac{r^2E}{f(r)},
    \label{t}
\end{equation}
\begin{equation}
    \frac{d\varphi}{d\eta}=\frac{L_z}{P(\theta)\sin^2{\theta}},
    \label{phi}
\end{equation}
\begin{equation}
   \left(\frac{dr}{d\eta}\right)^2=r^4 E^2-r^2f(r)\mathcal K,
   \label{r}
\end{equation}
\begin{equation}
    \left( \frac{d\theta}{d\eta}\right)^2=P(\theta)\mathcal{K}-\frac{L_z^2}{\sin^2{\theta}},
    \label{the}
\end{equation}
where $\eta$ is the Mino parameter related to the affine parameter $\lambda$ by
\begin{equation}
    \frac{d\eta}{d\lambda}=\frac{\Omega^2(r,\theta)}{r^2}.
\end{equation}
The integrals of motion $L_z$, $E$, and $\mathcal{K}$ arise when reducing the second order equations of motion to first order form. These constants represent the projected angular momentum ($L_z$), the particle energy ($E$), and the Carter constant ($\mathcal{K}$) \cite{Carter:1968rr}. Specifically, $\mathcal{K}$ generalizes the total angular momentum and emerges due to the separability of the $r$ and $\theta$ components. 

We are looking to cross the $\theta=\theta_+$ boundary, therefore we should look at what happens to \eqref{the} on this boundary. We notice that \eqref{the} becomes
\begin{equation}
    \left(\frac{d\theta}{d\eta}\right)^2=-\frac{L_z^2}{\sin^2\theta},\quad (\theta\to \theta_+).
\end{equation}
We conclude that a light ray with non zero angular momentum $(L_z\neq0)$ cannot ever get close to $\theta=\theta_+$, while if $L_z=0$, its $\theta$ velocity becomes zero there. Consequently one would need either one of two things so that photons would trespass this boundary: an imaginary angular momentum $L_z$ which would result in $(d\theta /d\eta)^2>0$ on $\theta=\theta_+$ or a divergent Carter constant $\mathcal{K}\to\infty$, that would cancel the zero of $P(\theta_+)$. In fact, provided that $E\neq0$, if one considers a constant $r$ motion along the black hole horizon $r_H$, from \eqref{r}, we note that 
\begin{equation}
    \mathcal{K}=\frac{r^2E^2}{f(r)} \to \infty \quad (r\to r_H),
\end{equation}
which is quite an artificial thing to happen, thus we surprisingly conclude that this boundary is generically not traversable. %

A follow up question whose answer might already be guessed is what does the far away observer whose time runs by coordinate $t$ sees? On this case, the velocity $d\theta/dt$ is given by
\begin{equation}
    \left( \frac{d\theta}{dt}\right)^2=\left(\frac{f(r)}{r^2E} \right)^2\left( P(\theta) \mathcal{K}-\frac{L_z^2}{\sin^2\theta}\right).
\end{equation}
We infer that once more $d\theta/dt\to 0$ as $\theta\to\theta_+$. If one would like otherwise, taking $L_z$ a complex number could solve the problem, but now a divergent Carter constant cannot.
Therefore, we find that contrary to generic event horizons of black holes (e.g. the Schwarzschild $r=2M$ horizon) which are crossable in a finite value of affine parameter, we find that such is not possible for the boundary $\theta=\theta_+$. This is to be expected as the topology of this regime remains a deformed 2-sphere with the given boundary being one of its axis of symmetry. %

Before analyzing the $P(\theta)>0$ region, let us understand what problems would arise associated to allowing a negative $P(\theta)$ function. The first detail that catches the eye is that the signature of the metric changes \eqref{Metric}. Provided that $t$ and $r$ remain timelike, we would now have a spacelike $\theta$ and $\varphi$ such that the signature is now $(-+--)$. A natural consequence of this fact is that it is easy to construct closed timelike curves. Consider a curve of constant $t$, $r$ and $\theta$, and let the coordinate $\varphi$ start at some value $\varphi_0$ and end at $\varphi_0+2\pi C$. By construction, this curve is closed, to see that it is also timelike we need to find that the length squared of this curve is negative. It is given by
\begin{equation}
    \Delta s^2=g_{\varphi\varphi}\Delta\varphi^2=P(\theta)\left(2\pi C\frac{r\sin\theta }{\Omega(r,\theta)}\right)^2<0.
\end{equation}
Therefore we have found a closed and timelike curve, which can but does not necessarily have to be a geodesic.

If we further investigate the consequences of the metric signature, consider an observer that is both inside the $P(\theta)<0$ region and the black hole event horizon, where $t$ and $r$ swap signature. The outcome is that, by starting with the metric convention $(-+++)$, the signature of the spacetime is transformed into $(+---)$. Consequently, the physics inside this region is identical to the physics outside it except for the simple property that timelike curves now became spacelike and vice versa, essentially meaning that timelike curves are now traveling faster than the speed of light.

Let us now come back to the physically relevant region where $P(\theta)>0$. This and the $P(\theta)<0$ are now separated by $\theta=\theta_+$ region. Instinctively, this would seem that we could eliminate the conical singularities of the spacetime, since the conical singularity at $\theta=0$ is behind $\theta=\theta_+$. However, a new conical singularity at $\theta=\theta_+$ emerges. By performing the analysis of the conical defect at this region (as was done in \ref{Interpretation and source of acceleration}), we are computing the Circumference over $2\pi \times \text{Radius}$ of a circle approaching $\theta \to \theta_+$. Proceeding with the needed care, since the function $P(\theta)$ vanishes there we find that 
\begin{equation}
\lim_{\theta\to\theta_+}\frac{\text{Circumference}}{2\pi\times  \text{Radius}}=\lim_{\theta\to\theta_+}\frac{C\sqrt{P(\theta)}\sin \theta}{\int_{\theta_+}^\theta   1/\sqrt{P(\vartheta)}d\vartheta}=\frac{CP'(\theta_+)\sin \theta_+}{2}=\frac{C(r_+-r_-)(r_+^2-r^2_A)}{2r_+^2r_A},
\end{equation}
where the prime denotes derivative with respect to $\theta$.
For this reason, we can choose to remove this conical defect by imposing 
\begin{equation}
    C=\frac{2}{P'(\theta_+)\sin\theta_+}\eqcolon C_{\theta_+},
\end{equation}
but with the down effect that at $\theta=\pi$ there is a negative angle defect 
\begin{equation}
    \delta_\pi^{\theta_+}=2\pi \left(1-\frac{2P(\pi)}{P'(\theta_+)\sin\theta_+}\right)=-2\pi \frac{(r_A+r_+)(2r_+r_-+r_A(r_+-r_-))}{r_A(r_+-r_-)(r_+-r_A)}<0,
    \label{Super accelerating strut}
\end{equation}
signaling a strut.

Or equivalently, we can choose to remove the strut at $\theta=\pi$ by imposing $C=C_\pi$ given in \eqref{Cpi}, however introducing a deficit angle at $\theta=\theta_+$ given by
\begin{equation}
    \delta_{\theta_+}^\pi=2\pi \left(1-\frac{P'(\theta_+)\sin\theta_+}{2P(\pi)}\right)=2\pi\frac{(r_A+r_+)(2r_+r_-+r_A(r_+-r_-))}{2r_+^2(r_A+r_-)}>0,
    \label{delta thetap}
\end{equation}
signaling a string.


A question from section \ref{Common accelerating regime} regarding the condition $\delta_0^\pi<2\pi$ was left to answer here. The fact was that this condition is violated for the first super accelerating regime \eqref{delta0 condition}, however the answer should now be clear. The solution is that we are not allowed to get close to the string at $\theta=0$, therefore what we should evaluate is not $\delta_0^\pi<2\pi$, but rather $\delta_{\theta_+}^\pi<2\pi$. This imposes that 
\begin{equation}
\frac{P'(\theta_+)\sin\theta_+}{P(\pi)}>0\iff(r_+-r_A)(r_+-r_-)>0,
\end{equation}
which is always true in the super accelerating regime.


A following question which was left open in section \ref{Common accelerating regime} was regarding the strut configuration. This, on the first super accelerating regime would transform into a string which would accelerate the black hole in the opposite direction it was previously moving. We should remind ourselves that this configuration was present by a removal of the conical singularity at $\theta=0$. As we have done in \eqref{Super accelerating strut}, we should instead remove the conical singularity at $\theta=\theta_+$ and find that indeed, the strut does not change into a string, providing the solution to this issue. More generally, we have found that the strut at $\theta=\pi$ will remain a strut (provided that one removes the conical singularity at $\theta=\theta_+$) and there is now present a cosmic string at $\theta=\theta_+$ (provided that one removes the conical singularity at $\theta=\pi$).

Yet, the string that was present on $\theta=0$ is now in an inaccessible region. Even though it might be out of causal contact with the region $\theta\in(\theta_+,\pi)$, it is interesting to know whether it remains a string. Removing the conical singularity at $\theta=\pi$ gives \eqref{delta0} and thus $\delta_0^\pi>0$, which means that we continue having a string at $\theta=0$. Also, if we remove the conical singularity at $\theta=\theta_+$ we find that at $\theta=0$ there is a conical defect 
\begin{equation}    \delta_{0}^{\theta_+}=2\pi\left(1-\frac{2P(0)}{\sin{\theta_+}P'(\theta_+)}\right)=2\pi\frac{r_A^2(r_+-r_-)+r_+r_A(3r_+-r_-)-2r_-r_+^2}{(r_A+r_+)(r_+-r_-)r_A}.
\label{delta0thetap}
\end{equation}
This is always positive provided that $r_-<r_A<r_+$, signaling that the string at $\theta=0$ is preserved even by removing the conical singularity at $\theta=\theta_+$ and creating a strut at $\theta=\pi$. This would, therefore, create a spacetime with a total of $2$ conical singularities.


Now we are left to analyze conformal infinity. Once again, it is satisfied by the pair $(r,\theta)$ such that 
\begin{equation}
    r\cos\theta=r_A.
\end{equation}
We restrict $r\in (r_A,r_+)$ and $\theta\in(\theta_+,\pi)\iff\cos\theta\in(-1,r_A/r_+)$. We find that, conformal infinity is only reached for the pair $(r,\theta)=(r_+,\theta_+)$, which is a point where the angular and cosmological horizons of the black hole meet. 

If we would like to find out how conformal infinity behaves inside $\theta\in(0,\theta_+)$, we would see that it is a surface of constant height $z$ that intersects the event horizon of the black hole at $\theta=0$, and is generically described by $z=r\cos\theta=r_A$. 

Due to conformal infinity being reached on the cosmological horizon $r_+$, the topology of this surface is different from any other constant $r$ surface. Notice that on this region, a constant $r=r_+$, $t$ and $\varphi$ curve, as length given by \eqref{length theta} which diverges since both $P(\theta),\Omega(r,\theta)\to0$. However, to find out what happens to the length of a closed $\varphi$ curved of a constant $t$, $\theta$ and $r=r_+$, we should look at \eqref{length varphi} in the limit as $\theta \to \theta_+$, implying that
\begin{equation}
    \lim_{\theta\to \theta_+} \ell_\varphi(\theta)=2\pi Cr\lim_{\theta\to \theta_+}\frac{\sin \theta\sqrt{P(\theta)}}{1-r_+r_A^{-1}\cos\theta}=\infty,
    \label{limit length vaphi}
\end{equation}
where we have used the fact that $\theta_+$ is a simple zero of $P(\theta)$ as this would change the result in \eqref{limit length vaphi}\footnote{If $\theta_+=0$ the $\sin\theta$ function also goes to zero, but the end result remains divergent. 
The function $P(\theta)$ can indeed develop a degenerate zero in the second super accelerating regime if $r_+=r_-$, this will be analyzed in more detail in section \ref{Super accelerating Nariai}}. So we find once more that the topology of this surface is $\mathbb{R}_t\times \mathbb{R}^2=\mathbb{R}^3$.


One can visualize the structure of the first super accelerating regime on figure \ref{Figure first super accelerating regime}, however, this picture should be analyzed with care, as it might lead one to infer that the $\theta=\theta_+$ are cones, which we have alluded to the fact that they are not. Additionally, there is the fact that we are representing constant $r$ surfaces as spheres, which can be dangerous, because as we have seen, the black hole is not really spherically symmetric.

\FIGURE{
    \centering
    \includegraphics[width=0.6\linewidth]{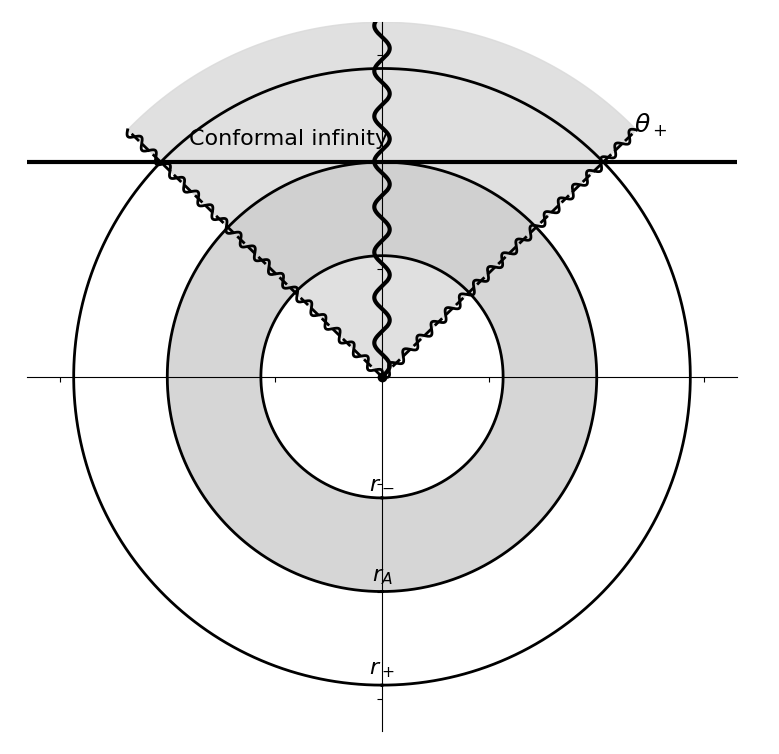}
    \caption{Two dimensional representation of the horizon structure of the C-metric on the first super accelerating regime. On the vertical axis we have the $z=r \cos\theta$ and on the horizon axis we have $x=r\sin \theta$. The horizons of the black hole are plotted and the shaded regions represent when a coordinate changes from the conventional signature (either $t,r$ or $\theta$). $r_-$ is the Cauchy horizon, $r_A$ the event horizon and $r_+$ the cosmological horizon. The new axis of symmetry is represented by the constant $\theta=\theta_+$ dashed and wavy line. Conformal infinity is represented by the constant $z=r_A$ line and the conical defect at $\theta=\pi$ was removed resulting in the cosmic string configurations at $\theta=\theta_+$ and $\theta=0$.}
    \label{Figure first super accelerating regime}}

\subsection{Second super accelerating regime ($r_A<r_-<r_+$)}\label{Second super accelerating regime}
\noindent
The second super accelerating regime is considered when the ascending order of the horizons of the black hole becomes $r_A<r_-<r_+$. This means that $r_A$ is the Cauchy horizon, $r_-$ the event horizon and $r_+$ the cosmological horizon of the black hole.

A new feature emerges on the angular sector. There is a second zero of the function $P(\theta)$ set by $\theta=\theta_-\coloneq \arccos{\left(\frac{r_A}{r_-}\right)}$. Therefore the $\theta$ and $\varphi$ coordinates become spacelike for $\theta\in (\theta_-,\theta_+)$. 
We are left to choose where to place the observer, either in $\theta \in (0,\theta_-)$ or $\theta \in (\theta_+,\pi)$ where in principle both configurations are allowed. Note that the maximum value that $\theta_\pm $ can attain is $\pi/2$, which would correspond to a divergent acceleration $A\to +\infty$. 

The discussion in \ref{First super accelerating regime} regarding the topology of the black hole and the conical singularities for $\theta \in (\theta_+,\pi)$ remains valid here. Moreover, we can also apply the same arguments to confirm that for $\theta\in (0,\theta_-)$ the angular coordinates also define a topology conformal to a deformed 2-sphere. 
Quantitatively however, the angle defect is now different at $\theta=\theta_-$. On the following, we will have a long discussion regarding all the conical singularities this regime can exhibit. The qualitative results are summarized on table \ref{Stringsandstrutstable}.

Consider a small circle close to $\theta=\theta_-$, its Circumference over $2\pi \times \text{Radius}$ is given by
\begin{equation}
\lim_{\theta\to\theta_-}\frac{C\sqrt{P(\theta)}\sin \theta}{\int_{\theta}^{\theta_-} 1/\sqrt{P(\vartheta)}d\vartheta}=-\frac{CP'(\theta_-)\sin\theta_-}{2}=\frac{C(r_+-r_-)(r_-^2-r_A^2)}{2r_-^2r_A},
\end{equation}
where the prime denotes derivative with respect to $\theta$. 

If one removes the strut at $\theta=\pi$, we find that, even though hidden behind the $\theta=\theta_+$ region, a resulting object on $\theta=\theta_-$ and $\theta=0$ is present, resulting in the total of $3$ conical singularities for the black hole. It is worth going through a simple calculation to see whether these are strings or struts. We perform identically as in \ref{First super accelerating regime} and \ref{Interpretation and source of acceleration} to find that if one removes the angle defect at $\theta=\pi$, there is a resulting angle deficit at $\theta=\theta_+$ given by \eqref{delta thetap} where $\delta_{\theta_+}^\pi>0$. At $\theta=\theta_-$,
\begin{equation}
    \delta_{\theta_-}^{\pi}=2\pi \left( 1+\frac{\sin\theta_- P'(\theta_-)}{2P(\pi)}\right)=2\pi \frac{r_A^2(r_+-r_-)+2r_-^2r_++r_Ar_-(3r_--r_+)}{2r_-^2(r_A+r_+)}>0
\end{equation}
and at $\theta=0$, $\delta_0^\pi>0$ is given by \eqref{delta0}. So it results that all these defects are represented by cosmic strings which push the black hole to the north direction.

If one removes the conical singularity at $\theta_+$ we find that at $\theta=\pi$ there is a strut whose angle defect is given by \eqref{Super accelerating strut}. At $\theta=\theta_-$ we find that

\begin{equation}
    \delta_{\theta_-}^{\theta_+}=2\pi \left(1+\frac{\sin \theta_- P'(\theta_-)}{\sin\theta_+P'(\theta_+)}\right)=2\pi \left(2-\left(\frac{r_+-r_-}{r_A}\right)^2+\frac{(r_+-2r_-)r_+}{r_-^2}\right),
\end{equation}
which can either be a string or strut depending on the values of $r_\pm$ and $r_A$. 
Finally, at $\theta=0$ we find that the angle defect was calculated in \eqref{delta0thetap}, which can be positive of negative depending on $r_\pm$ and $r_A$, given that $r_A<r_-$. Consequently, we conclude that removing the conical singularity at $\theta=\theta_+$, while introduces a strut at $\theta=\pi$, the conical defect at $\theta=\theta_-$ and $\theta=0$ are not clearly either just strings or struts for the whole parameter configurations which define the second super accelerating regime.

For the purpose of removing the conical defect at $\theta_-$, one should choose
\begin{equation}
    C=\frac{-2}{P'(\theta_-)\sin\theta_-}\eqcolon C_{\theta_-}.
\end{equation}
Proceeding with this option introduces an angle defect at $\theta=0$ given by
\begin{equation}
    \delta_0^{\theta_-}=2\pi \left( 1+\frac{2P(0)}{P'(\theta_-)\sin\theta_-}\right)=-2\pi \frac{(r_--r_A)(r_A(r_+-r_-) +2r_+r_-)}{r_A(r_+-r_-)(r_A+r_-)}<0,
    \label{Defect at 0 Second regime}
\end{equation}
which shows that there is a strut there. At $\theta=\theta_+$ we find 
\begin{equation}
    \delta_{\theta_+}^{\theta_-}=2\pi \left( 1+\frac{\sin \theta_+P'(\theta_+)}{\sin\theta_-P'(\theta_-)}\right)=1+\frac{r_A^2r_-^2}{(r_+-r_-)^2(r_A^2-r_-^2)},
\end{equation}
which has a non trivial sign, depending on the specific values of $r_\pm$ and $r_A$.
At $\theta=\pi$ we notice
\begin{equation}
    \delta_{\pi}^{\theta_-}=2\pi \left(1+\frac{2P(\pi)}{\sin \theta_- P'(\theta_-)} \right)=-2\pi \frac{r_A^2(r_+-r_-)+r_-(2r_+r_-+3r_Ar_--r_Ar_+)}{r_A(r_--r_A)(r_+-r_A)}<0,
\end{equation}
indicating that a strut lives there. Therefore we conclude that removing the conical singularity at $\theta=\theta_-$ introduces a strut at $\theta=0$ and $\theta=\pi$, plus another defect at $\theta=\theta_+$. This is intriguing, as it could indicate that the black hole should be accelerating in the opposite direction we have described in \ref{Interpretation and source of acceleration}, now to the south $\theta=\pi$. On the other hand, that interpretation is open to debate, as it is not clear whether conical defects present behind $\theta=\theta_\pm$ should contribute to the source of acceleration of the black hole.

At last, we consider the final possible set of string/strut scenarios. We remove the conical defect at $\theta=0$ and notice that at $\theta=\theta_-$ there is an angle defect
\begin{equation}
    \delta_{\theta_-}^0=2\pi \left(1+ \frac{P'(\theta_-)\sin\theta_-}{2P(0)}\right)=2\pi\frac{(r_--r_A)(r_A(r_+-r_-)+2r_+r_-r_A)}{2r_-^2(r_+-r_A)}>0,
    \label{Defect at theta- Second regime}
\end{equation}
signaling that a string is present there. At $\theta=\theta_+$ 
\begin{equation}
    \delta_{\theta_+}^0=2\pi \left(1-\frac{\sin\theta_+P'(\theta_+)}{2P(0)} \right)=-2\pi\frac{r_A(r_+-r_A)(r_+-r_-)+2r_+^2(r_A+r_-)}{2r_+^2(r_--r_A)}<0,
\end{equation}
points out that a strut lives there. Finally, at $\theta=\pi$, the angle defect is given by \eqref{deltapi} with $\delta_\pi^0<0$, showing a strut. Accordingly, removing the conical singularity at $\theta=0$ leaves us with the curious structure of a string at $\theta=\theta_-$ and struts at $\theta=\theta_+$ and $\theta=\pi$.

A summary of the resulting string/strut configurations is given in
table \ref{Stringsandstrutstable}.
\begin{table}[htbp]
    \centering
    \renewcommand{\arraystretch}{1.2}
    \begin{tabular}{c|cccc}
        \toprule
        \multirow{2}{*}{Regularized axis}
        & \multicolumn{4}{c}{Resulting configuration} \\
        \cmidrule(lr){2-5}
        & $\theta=0$ & $\theta=\theta_-$
        & $\theta=\theta_+$ & $\theta=\pi$ \\
        \midrule
        $\theta=0$       & Regular & String & Strut & Strut \\
        $\theta=\theta_-$ & Strut & Regular & ? & Strut \\
        $\theta=\theta_+$ & ? & ? & Regular & Strut \\
        $\theta=\pi$     & String & String & String & Regular \\
        \bottomrule
    \end{tabular}
    \caption{String/strut configurations obtained by removing the
    conical defect along the indicated axis of symmetry. ``Regular''
    denotes the absence of a conical defect, while ``String'' and
    ``Strut'' denote the corresponding configurations. A question mark
    indicates that the configuration is parameter-dependent and may
    correspond to either a string or a strut, depending on the black
    hole's mass, charge and acceleration.}
    \label{Stringsandstrutstable}
\end{table}

Once more, it is not entirely clear whether acceleration is provided by only one conical singularity or the effective contribution of all of them, as some are hidden inside inaccessible regions whose boundaries are set by $\theta=\theta_\pm$. As a consequence, it could potentially change the interpretation of each configurations completely. Either way, we have seen that possibly the best well behaved configuration is the one where $\theta \in (\theta_+,\pi)$ by the removal of the conical singularity at $\theta=\pi$, due to the clear interpretation that acceleration is provided to the north direction. Further negative comments on the regime $\theta\in (0,\theta_-)$ will be discussed on section \eqref{Geometrical and physical quantities}.

Focusing on conformal infinity, we once more notice that it is reached by
\begin{equation}
    r\cos\theta=r_A.
\end{equation}
Remember that $r\in (r_-,r_+)$ and let us first consider that we live on $\theta\in (\theta_+,\pi)$. On this range of $\theta$, we find that conformal infinity is placed on the same region as in the first super accelerating regime (discussed in section \ref{First super accelerating regime}), which is the pair $(r,\theta)=(r_+,\theta_+)$, obtained on the cosmological horizons of the black hole. On the other hand, should we consider $\theta\in (0,\theta_-)$, we find that conformal infinity is obtained for timelike $(r,\theta)=(r_-,\theta_-)$ which marks the event horizons of the black hole. This fact, therefore, will change the topology of the event horizon of this configuration.

Hence, from all configurations we have seen so far, we may conclude that conformal infinity is not reached in any region where the metric preserves the conventional $(-+++)$ signature. Moreover, for timelike $r$, we concluded that it stands in the middle of the $\theta_-$ and $\theta_+$ and therefore divides them into two separated spacetime regions, as described right before subsection \ref{Horizon structure}.

To examine the topology of the cosmological horizon, we perform a similar analysis as in section \eqref{First super accelerating regime}. We once again find that due to conformal infinity being reached, the topology of the cosmological horizon set by $r=r_+$ when $\theta \in(\theta_+,\pi)$ is $\mathbb{R}_t\times \mathbb{R}^2=\mathbb{R}^3$. However, when $\theta\in(0,\theta_-)$ it is the event horizon $r_A$ which has the topology of $\mathbb{R}_t\times \mathbb{R}^2=\mathbb{R}^3$. All the other constant $r$ surfaces have the usual time $\mathbb{R}_t$ times another deformed 2-sphere $\bar S^2$ topology.

On figure \ref{Figure second super accelerating regime}, we see a visual representation of the structures of the black hole we have been considering. Once more, we remind the reader that this should be viewed with care, since the $\theta=\theta_\pm$ regions, are represented by cones but in fact are just new axes of symmetry of the angular manifold. Moreover, we are representing constant $r$ surfaces by spheres, which can be misleading, as the black hole is not spherically symmetric. Additionally, on figure \eqref{Figure second super accelerating regime} we have decided to remove the strut at $\theta=\pi$ resulting in cosmic strings on $\theta=\theta_\pm$ and $\theta=0$.

\FIGURE{
    \centering
    \includegraphics[width=0.6\linewidth]{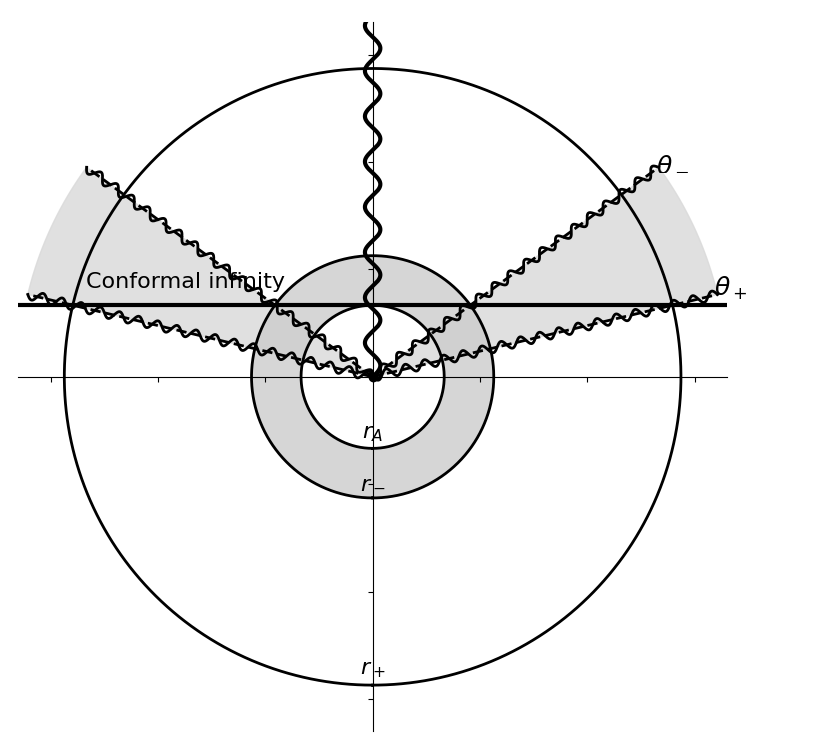}
    \caption{Two dimensional representation of the horizon structure of the C-metric on the second super accelerating regime. On the vertical axis we have the $z=r \cos\theta$ and on the horizon axis we have $x=r\sin \theta$. The horizons of the black hole are plotted and the shaded regions represent when a coordinate changes from the conventional signature (either $t,r$ or $\theta$). $r_A$ is the Cauchy horizon, $r_-$ the event horizon and $r_+$ the cosmological horizon. The new axis of symmetry are represented by constant $\theta=\theta_+$ and $\theta=\theta_-$ dashed and wavy lines. Conformal infinity is represented by the constant $z=r_A$ line and the conical defect removed is at $\theta = \pi$ resulting in cosmic strings at $\theta=\theta_\pm$ and $\theta=0$, represented by the wavy lines.}
    \label{Figure second super accelerating regime}}

\subsection{Super accelerating regime with a cosmological constant}\label{Super accelerating regime with cosmological constant}
\noindent
We now consider a non zero cosmological constant. This does not affect the function $P(\theta)$, nor $\Omega(r,\theta)$, therefore the bulk of the discussion regarding conformal infinity, and the string/strut picture done in the three previous sections remains. What we wish to explore here are the new horizon possibilities that emerge. Let us consider the function $f(r)$ given by \eqref{metric functions} by setting $\Lambda=0$ and $\tilde f(r)=f(r)-\Lambda r^2/3$ or $\tilde f(r)=f(r)+r^2/L^2$ (in case $\Lambda<0$) which are identical to $f(r)$ in \eqref{metric functions}.
This addition of the cosmological constant allows us to construct exotic configurations such as naked singularities or even horizonless objects in AdS space.

Let us start by reviewing a well known property of the C-metric in AdS \cite{Dias:2002mi}. Here, it is possible to eliminate the cosmological horizon which previously was set by $r_A$. To do so, notice that 
\begin{equation}
\tilde f(r)=1-\frac{2M}{r}+\frac{Q^2}{r^2}-A^2Q^2+2A^2Mr+\left(\frac{1}{L^2}-A^2\right)r^2.
\end{equation}
If $A=1/L$, the repulsion of acceleration $A$ and $1/L^2$ exactly cancel each other out and the cosmological horizon disappears. If $A>1/L$ the cosmological remains, but if $A<1/L$ the spacetime becomes asymptotically AdS eliminating the cosmological horizon of the spacetime. 

To construct a naked singularity we need to be in AdS as $\tilde f(r)>f(r)$. To find out which parameters of $A,M,Q$ lead to this configuration consider the following construction. We take the zeros $r_A$ and $r_-$ of the function $f(r)$ and impose that we are close to the transition from the first super accelerating regime to the second one, by restricting $r_A\sim r_-$. Whereas $f(r)$ is close, but less than zero ($f(r)\lesssim 0$) in the region $r\sim r_A, r_-$, the new horizon function on the same region $\tilde f(r)=f(r)+r^2/L^2\sim r^2/L^2$ is now positive. This means the that horizons $r_A$ and $r_-$ were eliminated and we end up with a naked singularity. The precise range of values $M,Q$ and $A$ needed to exhibit this configuration is difficult to express analytically; yet, with this argument, we know that the closer $r_A$ is to $r_-$, the more relaxed are the upper bounds on the AdS length $L$. As $r_A$ starts to distance from $r_-$, $L$ should become smaller so as to preserve the naked singularity. Furthermore, this naked singularity configuration can additionally be divided into two distinct spacetimes. One of such is a naked singularity with a cosmological horizon, as long as $L>r_A$; the other is an horizonless configuration which is present when $L\leq r_A$.

\section{Geometrical and physical properties}  \label{Geometrical and physical quantities}
\noindent
Having discussed the super accelerating regimes, we now turn our attention to the parameters $M$ and $Q$. Once more, the cosmological constant is turned off by setting $\Lambda=0$. From works on the C-metric \cite{Griffiths:2006tk}, it is known that the mass and charge of the black hole are not simply the parameters $M$ and $Q$. One usually uses the concept of Komar mass and charge to compute the actual mass and charge of the black hole. 
Let us focus our attention on the Komar mass, where for a charged black hole is defined as 
\begin{equation}
    M_{K}=-\frac{1}{8\pi}\int \nabla ^\mu K^\nu dS_{\mu \nu}+\frac{1}{4\pi}\int K^\mu A_\mu F^{\alpha \beta}dS_{\alpha \beta}.
    \label{Komar Mass I}
\end{equation}
$K^\mu =(\partial_t)^\mu$ is the time like Killing vector associated with time translation symmetry, $dS_{\mu \nu}$ is the infinitesimal area surface vector and $F_{\alpha\beta}=\partial_\alpha A_\beta -\partial_\beta A_\alpha$ is the electromagnetic field strength tensor.

A known property of the Komar mass integral \eqref{Komar Mass I} is that it yields the same result when evaluated on any two homologous closed compact orientable surfaces which enclose the matter content of the spacetime\footnote{Two surfaces are homologous if one can be continuously deformed into the other through without crossing any obstruction, singularity or boundary.}, \cite{Natario:2020nvk}. On the following, we will compute this integral along a constant $r$ and varying $\theta,\varphi$ spatial surfaces for both the regular and super accelerating regimes. On the super accelerating regimes we are not allowed to integral over the whole range of $\theta\in(0,\pi)$, but instead on a smaller subset. However, as we have allude before in \ref{First super accelerating regime}, we will integrate over a surface which is topologically a deformed 2-sphere, and therefore, the surface we are integrating over is closed. Either way, we can nevertheless perform the whole calculation for a generic $r$ and find out that the obtained result will be $r$ independent. As such, the Komar mass in the super accelerating regime is reasonably well defined, apart from an ambiguity that also arises in the regular accelerating regime: the normalization of the time translation Killing vector, since the spacetime does not become flat as $r\to+\infty$. We leave this ambiguity unresolved.

Its calculation will be performed now. For static spacetimes where the metric is written in the form
\begin{equation}
    ds^2=-N^2dt^2 +h_{ij}dx^idx^j,
\end{equation}
the first term on the right hand side of \eqref{Komar Mass I} can be written as
\begin{equation}
    \frac{1}{4\pi}\int n^i \partial_i Nd\mathcal{A},
    \label{Komar Mass II}
\end{equation}
where $n^i$ is the unit vector normal to the induced surface and $d\mathcal{A}= \sqrt{g_{\theta \theta}g_{\varphi \varphi}}d\theta d\varphi$ is the area element of such surface whose total value is 
 \begin{equation}
     \mathcal{A}=\int \sqrt{g_{\theta \theta}g_{\varphi \varphi}}d\theta d\varphi=2\pi Cr^2\int \frac{\sin\theta}{(1-Ar\cos\theta)^2}d\theta.
     \label{Area}
    \end{equation}

Applying this to a constant $t$ and $r$ surface where $N=\sqrt{f}/\Omega$ and $n^i=\Omega\sqrt{f}\delta^i_r$, equation
\eqref{Komar Mass II} is written as
\begin{equation}
    \int\left[\frac{f'(r)}{8\pi}-\frac{f(r)}{4\pi} \frac{\partial_r\Omega(r,\theta)}{\Omega(r,\theta)}\right]d\mathcal{A}=\frac{f'(r)}{8\pi}\mathcal{A}+\frac{f(r)r^2}{2}C\int \frac{A\cos\theta \sin \theta}{(1-Ar\cos\theta)^3}d\theta.
    \label{Komar Mass III}
\end{equation}
On the other hand, the second term in \eqref{Komar Mass I} can be computed easily if we understand that $dS_{rt}=d\mathcal{A}/\Omega^2$, due to the unit vectors normal to the surface of constant $t$ and $r$.

Using the electric potential \(\mathscr{A}_t=-Q/r\) and the only non vanishing component of the Faraday tensor, \(F_{rt}=Q/r^2\), the second term in \eqref{Komar Mass I} becomes
\begin{equation}
    \frac{Q^2C}{2r}\int \sin\theta d\theta.
    \label{Komar Mass IV}
\end{equation}
Thus the resulting Komar mass of the C-metric is given by
\begin{equation}
    M_K=\frac{f'(r)}{8\pi}\mathcal{A}+\frac{f(r)r^2}{2}C\int \frac{A\cos\theta \sin \theta}{(1-Ar\cos\theta)^3}d\theta+\frac{Q^2C}{2r}\int \sin\theta d\theta.
    \label{Komar Mass Final}
\end{equation}

Now it is time to perform the integrals. On the regular accelerating regime $\theta\in (0,\pi)$, we find that \eqref{Komar Mass Final} becomes $r$ independent as expected and it is given by
\begin{equation}
    M_K=CM,
    \label{Komar mass common accelerating regime}
\end{equation}
showing that the Komar mass not only depends on the mass parameter $M$, but also on the charge and acceleration parameters present on the $C$ term which is defined depending on the conical singularity one wishes to remove from the spacetime, or else, if one chooses to not remove any of the singularities, the mass parameter and the Komar mass become identical.

On the first super accelerating regime, the integration is performed in the range $\theta\in(\theta_+,\pi)$ and we also find that the $r$ dependence of the Komar mass \eqref{Komar Mass Final} is gone. Not only that, but the expression obtained gives
\begin{equation}
    M_K=\frac{C}{4A}\left(1+2MA+A^2Q^2\right).
    \label{Komar mass first super accelerating regime}
\end{equation}
We have argued previously \ref{First super accelerating regime} that the most reasonable configuration is one where there in no conical defect at $\theta=\pi$ and the string at $\theta=0$ is hidden. This corresponds to choosing $C=C_\pi$ defined in \eqref{Cpi}. Given that, we miraculously find that 
\begin{equation}
    M_K=\frac{1}{4A}.
    \label{1/4A}
\end{equation}
Consequently, the Komar mass of the C-metric in the super-accelerating regime is determined solely by the acceleration parameter, up to the normalization of the time translation Killing vector.

Proceeding to the second super accelerating regime we find two options. One is that $\theta\in(\theta_+,\pi)$ which recovers \eqref{Komar mass first super accelerating regime} and consequently, by effectively removing the $\theta=\pi$ singularity, we obtain \eqref{1/4A}. 
If we consider the region $\theta\in(0,\theta_-)$ we find that the $r$ dependence in the Komar mass \eqref{Komar Mass Final} is gone and it gives
\begin{equation}
    M_K=-\frac{C}{4A}\left(1-2MA +A^2Q^2\right).
\end{equation}
Curiously, the result is negative. 
This might be an indication to discard that $\theta$ regime, but an interpretation to why it is so remains unknown.
If we ignore the fact that $M_K<0$ and we wish to remove the conical singularity at $\theta=0$ of the configuration in question as described in \ref{Second super accelerating regime}, we find that $M_K$ simplifies nicely as before, but now with a negative sign
\begin{equation}
    M_K=-\frac{1}{4A}.
\end{equation}
To the Komar mass, this is all we had to present, although this brings some open new questions which to be discussed on \ref{Conclusion}.

Let us focus our attention on the charge of the black hole. The discussion is much simpler, we use Gauss's law to define the Komar charge as 
\begin{equation}
    Q_K=\frac{1}{4\pi}\int F^{\mu \nu}dS_{\mu \nu}=\frac{QC}{2}\int \sin \theta d\theta.
\end{equation}
Then it is trivial to obtain that on the regular accelerating regime 
\begin{equation}
    Q_K=CQ,
    \label{Komar charge common regime}
\end{equation}
whose form is quite similar to \eqref{Komar mass first super accelerating regime}. On the first and second super accelerating regime, when $\theta\in (\theta_+,\pi)$, we find that 
\begin{equation}
    Q_K=\frac{CQ(r_++r_A)}{2r_+},
    \label{Komar charge super acclerating}
\end{equation}
which in general is less than \ref{Komar charge common regime}. 

On the other hand if we consider the configuration in $\theta \in (0,\theta_-)$ we find that 
\begin{equation}
    Q_K=\frac{CQ(r_--r_A)}{2r_-}.
\end{equation}
We now find that in contrast to the Komar mass, the Komar charge does not change sign on this configuration. 

We are left to discuss the area $\mathcal{A}$ of the event and cosmological horizons. We have mentioned before that the area of a constant $r$ surface\footnote{We are omitting time for simplicity, every time we mention a constant $r$ surface we are also implicitly imposing constant $t$ as well.} is given by \eqref{Area} where now we shall evaluate it at the corresponding coordinate $r$ representing the cosmological and event horizons.

On the regular accelerating regime where the event horizon is defined by $r_+$, it is well known that 
\begin{equation}
    \mathcal{A}=\frac{4\pi Cr_+^2r_A^2}{r_A^2-r_+^2}.
\end{equation}
On the other hand, the literature is also familiar with the property that the area of the cosmological horizon defined by the constant $r=r_A$ surface is divergent. The reason for this we have is that conformal infinity is reached when the integrand reaches $\theta\to0$ and thus changes the topology of this horizon, making it a non compact surface.

On the first super accelerating regime, the event horizon is established by the surface $r=r_A$ and we shall integrate over $\theta \in (\theta_+,\pi)$. Therefore the event horizon area becomes
\begin{equation}
    \mathcal{A}=\frac{\pi Cr_A^2(r_A+r_+)}{r_+-r_A}.
\end{equation}
If we wish to compute the area of the cosmological horizon, we find that it diverges once more, and the reason is again due to conformal infinity being reached as $\theta\to\theta_+$, when $r=r_+$.

Let us proceed to the second super accelerating regime, where the coordinate $r_-$ represents now the event horizon of the black hole. If we are now limited by $\theta\in(\theta_+,\pi)$, the area of the event horizon is
\begin{equation}
    \mathcal{A}=\frac{2\pi Cr_-^2 r_A(r_A+r_+)}{(r_+-r_-)(r_-+r_A)}.
\end{equation}
Regarding the area of the cosmological horizon, the story is the same, it diverges once more due to conformal infinity.

We come to analyze the last possible case. The second super accelerating regime with $\theta \in(0,\theta_-)$. Notice that conformal infinity is now reached on the event horizon of the black hole plus when $\theta\to \theta_-$, therefore we find that it is the area of the event horizon which is divergent.
So we are left to find the only possibility where the area of the cosmological horizon is finite and it is given by
\begin{equation}
    \mathcal{A}=\frac{2\pi Cr_+^2r_A(r_A-r_-)}{(r_+-r_-)(r_A-r_+)}.
\end{equation}

\section{Near extremal limits}\label{Near extremal}
\noindent
In this section, we investigate the near extremal configurations of the C-metric. This has been first studied by Dias and Lemos in \cite{Dias:2003up}, but using a different coordinate system and the super accelerating regime had not been identified.
Since the C-metric has three real horizons, its near extremal limits can be put into two categories: the Nariai configuration where the Cosmological horizon and the event horizon of the black hole are very close to each other, and the Bertotti-Robinson configuration where the Cauchy horizon and the event horizon of the black hole meet. As we have been discussing, the super accelerating regime swaps the character of the $r_\pm$ and $r_A$ horizons, therefore we are able to find super accelerating near extremal configurations. On the following subsections we will classify these regimes and find out their near horizon geometry. A non zero cosmological constant will be considered for simplicity, but results generalize in a straightforward, but tedious way. In the end, figure \ref{Phase space of extremal configurations} shows the parameter space which allow these extremal configurations.

\subsection{Accelerating Bertotti-Robinson} \label{Accelerating Bertotti-Robinson}
\noindent
We start by defining the accelerating Bertotti-Robinson configuration as a regular accelerating regime when the Cauchy horizon is close to the event horizon, such that $r_-\sim r_+<r_A$.

To investigate the near horizon limit, we impose that the two horizons are related by
\begin{equation}
    r_-=\rho -\epsilon, \quad r_+=\rho+\epsilon \quad \text{with} \quad 0<\epsilon \ll 1,
\end{equation}
where $\rho=M$ is the value of the $r$ coordinate in which the horizons would merge and $\epsilon$ is a small parameter that quantifies how far apart the two horizons are.

We now zoom into the horizon by introducing the coordinates such that
\begin{equation}
    t=\frac{\tau}{\epsilon\mathcal{K}},\quad r=\rho +\epsilon\cosh{\upsilon},
\end{equation}
where 
\begin{equation}
    \mathcal{K}=\frac{1-A^2\rho^2}{\rho^2}.
\end{equation}
Now by taking the limit $\epsilon \to0$ we find that the metric becomes
\begin{equation}
    ds^2= \frac{r_A^2}{(r_A-\rho\cos{\theta})^2}\left( \frac{1}{\mathcal{K}}\left(-\sinh^2\upsilon d\tau^2+d\upsilon^2\right)+\frac{\rho^2}{P(\theta)}d\theta^2+\rho^2P(\theta)\sin^2\theta d\theta^2\right).
    \label{Accelerating Bertotti-Robinson metric}
\end{equation}
Additionally, the electromagnetic field strength tensor becomes
\begin{equation}
    F=\frac{1}{\rho\mathcal{K}}\sinh \upsilon d\upsilon \wedge d\tau.
\end{equation}
Analyzing the geometry realized by \eqref{Accelerating Bertotti-Robinson metric} we find that the resulting spacetime is topologically conformal to $AdS^2\times \tilde S^2$, where $\tilde S^2$ represents a deformed 2-sphere due to the lack of spherical symmetry introduced by the $P(\theta)$ term. At every point in the deformed 2-sphere corresponds an anti de Sitter ($AdS^2$) spacetime except for a single point corresponding to an infinite string or strut discussed in section \ref{Horizon structure}.

\subsection{Accelerating Nariai}\label{Acclerating Nariai}
\noindent
The accelerating Nariai spacetime is defined by the regular accelerating regime when the event horizon approximates the cosmological horizon, such that $r_-<r_+\sim r_A$.

However, compared to case considered in the previous section \ref{Accelerating Bertotti-Robinson} a new problem emerges. While varying the parameters $M$ and $Q$ will simultaneously change the value of $r_\pm$, the values of $A$ do not affect the horizon $r_\pm$ at the same time that the values of $M,Q$ do not influence the horizon $r_A$. It is important that they do, since we have to require the following argument. There should be a parameter $\epsilon$ and a fixed $\rho$, such that
\begin{equation}
    r_A=\rho+\epsilon, \quad r_+=\rho-\epsilon \quad \text{with }|\epsilon|\ll1.
    \label{Nariaiac}
\end{equation}
Applying this construction therefore introduces an artificial constraint since generally, the two horizons do not need to be spaced by the same $\epsilon$ value. There could be another small number $\delta$ such that while $r_+=\rho -\epsilon$, $r_A=\rho+\delta$. Either way, we will proceed with the construction \eqref{Nariaiac}, by defining $\rho=(r_A+r_+)/2$. There is one logical way to justify this. Consider that the cosmological constant $\Lambda\neq0$; then we would find that all the black hole horizons actually depend on all its parameters, $M,Q$ and $A$, not forgetting $\Lambda$. Consequently, introducing the same displacement $\epsilon$ would not add any artificial constraint. As such, one can think of \eqref{Nariaiac} as arising naturally in the limit of small cosmological constant, as $\Lambda\to0$. Moreover, notice that we do not need $\epsilon>0$, since the near extremal limit can be taken either from the super or regular accelerating regimes. 

We can proceed to investigate the near horizon geometry by introducing the coordinate transformation
\begin{equation}
   t=\frac{\tau}{\epsilon\mathcal{K}},\quad r=\rho +\epsilon\cos{\upsilon},
\end{equation}
with
\begin{equation}
    \mathcal{K}=2\frac{\rho-r_-}{\rho^3}.
\end{equation}
Then, by taking the limit $\epsilon\to0$ we obtain the metric
\begin{equation}    
    ds^2=\frac{r_A^2}{(r_A-\rho\cos{\theta})^2}\left( \frac{1}{\mathcal{K}}\left( -\sin^2\upsilon d\tau^2+d\upsilon^2\right)+\frac{\rho^2}{P(\theta)}d\theta^2+\rho^2P(\theta)\sin^2\theta d\varphi^2\right)
    \label{Accelerating Nariai metric}
\end{equation}
and the electromagnetic field strength tensor
\begin{equation}
    F=-\frac{Q}{\rho^2\mathcal{K}}\sin{\upsilon} d\upsilon\wedge d\tau.
\end{equation}

Investigating the geometry induced by \eqref{Accelerating Nariai metric}, we conclude that the topology of this spacetime is $dS^2\times \tilde S^2$. At each point on the (distorted) 2-sphere corresponds a 2 dimension de Sitter spacetime ($dS^2$), except one special unavoidable point corresponding to a string or strut extending to infinity.
In \cite{Luo:2024cwm}, and in \ref{Common accelerating regime} we have argued that the topology of the acceleration horizon on a constant time slice, is $\mathbb{R}_t\times \mathbb{R}^2=\mathbb{R}^3$ instead of $\mathbb{R}\times \tilde S^2$ due to non compactness introduced by conformal infinity. 
Even though we indeed made the two horizons stick close together, we are analyzing the near, but not actual extremal case which corresponds to the equality $r_A=r_+=\rho$ and would thus impose the topology $\mathbb{R}^3$. The same argument will apply for subsection \ref{Super accelerating Bertotti-Robinson}.

\subsection{Super accelerating Bertotti-Robinson} \label{Super accelerating Bertotti-Robinson}
\noindent
The super accelerating Bertotti-Robinson spacetime is obtained in the super accelerating regime when the Cauchy horizon approaches the event horizon, such that $r_-\sim r_A <r_+$. Just like what happened in \ref{Acclerating Nariai}, the dependence of the black hole parameters in $r_-$ and $r_A$ splits, so the discussion there can be applied here.

Therefore define $\rho=(r_A+r_-)/2$ and consider that 
\begin{equation}
    r_A=\rho-\epsilon\quad \text{and}\quad r_-=\rho +\epsilon, \quad |\epsilon| \ll1.
\end{equation}
The quantity $\epsilon$ can equivalently be greater or smaller than zero meaning that this limit be obtained either from the first or second super accelerating regime.

To inquire on the near horizon geometry, we introduce the by now usual coordinate transformation
\begin{equation}
   t=\frac{\tau}{\epsilon\mathcal{K}},\quad r=\rho +\epsilon\cosh{\upsilon},
\end{equation}
with
\begin{equation}
    \mathcal{K}=2\frac{r_+-\rho}{\rho^3}.
\end{equation}
Taking the $\epsilon\to 0$ limit we find that the metric is
\begin{equation}
  ds^2=\frac{r_A^2}{(r_A-\rho\cos{\theta})^2}\left( \frac{1}{\mathcal{K}}\left(-\sinh^2{\upsilon }d\tau^2+d\upsilon^2\right)+\frac{\rho^2}{P(\theta)}d\theta^2+\rho^2P(\theta)\sin^2\theta d\varphi ^2\right),  
  \label{Super accelerating Bertotti-Robinson metric}
\end{equation}
while the electromagnetic strength tensor is
\begin{equation}
    F=\frac{Q}{\rho^2\mathcal{K}}\sinh{\upsilon} d\upsilon\wedge d\tau.
\end{equation}
Examining the geometry realized by \eqref{Super accelerating Bertotti-Robinson metric}, we find that the topology of this spacetime is once again conformal to $AdS^2\times \bar S^2$. By $\bar S^2$ we define another deformed 2-sphere whose topology was discussed in \ref{First super accelerating regime}. Thus, at every point on the deformed 2-sphere corresponds a 2 dimension de anti de Sitter spacetime ($AdS^2$), except one special unavoidable point corresponding to a string or strut extending to infinity.

\subsection{Super accelerating Nariai}\label{Super accelerating Nariai}
\noindent
The super accelerating Nariai results from the second super accelerating regime when the event horizon meets the cosmological horizon, such that $r_A<r_-\sim r_+$.

To analyze the near horizon geometry, we define $\rho=M$ as the coordinate on which the actual extremal configuration occurs. Further, we impose that the two horizons are related by a small parameter $\epsilon$ such that
\begin{equation}
    r_-=\rho -\epsilon ,\quad r_+=\rho + \epsilon, \quad \text{with } 0<\epsilon\ll 1.
\end{equation}

We then zoom into the horizons by performing the change of coordinates
\begin{equation}
    t=\frac{\tau}{\epsilon\mathcal{K}},\quad r=\rho +\epsilon\cos{\upsilon},
    \label{radial time zoom in super nariai}
\end{equation}
where 
\begin{equation}
    \mathcal{K}=\frac{A^2\rho^2-1}{\rho^2}.
    \label{K Nariai}
\end{equation}
As a result, when the limit $\epsilon\to 0$ is taken, the metric of the black hole becomes 
\begin{equation}
    ds^2=\frac{r_A^2}{(r_A-\rho \cos{\theta})^2}\left( \frac{1}{\mathcal{K}}\left( -\sin^2\upsilon d\tau^2+d\upsilon^2\right)+\frac{\rho^2}{P(\theta)}d\theta^2+\rho^2P(\theta)\sin^2\theta d\varphi^2\right)
    \label{super accelerating nariai metric}
\end{equation} 
while the electromagnetic field strength tensor is
\begin{equation}
    F=-\frac{1}{\rho\mathcal{K}}\sin \upsilon d\upsilon \wedge d\tau.
\end{equation}
Therefore, we can conclude that the topology of this spacetime is conformal to $dS^2\times \bar S^2$. At every spacetime point on the deformed 2-sphere will correspond a 2 dimensional de Sitter spacetime $dS^2$, expect for one special unavoidable point corresponding to a string or strut. 

So far, we have analyzed the $t,r$ components of the metric in the Nariai limit. However, we can also find that in this specific limit, $\theta_+\sim \theta_-$. So reverting \eqref{super accelerating nariai metric} back to the original metric \eqref{Metric}, we are ready to create a new that changes the angular sector of the metric. Define $\theta_0\coloneq \theta_+=\theta_-=\arccos(\frac{r_A}{\rho})$ as the angle at which the horizons merge, where $\rho=M$. Then consider that
\begin{equation}
    \theta_-=\theta_0 -\delta, \quad \theta_+=\theta_0 +\delta, \quad \text{with }0<\delta \ll 1. 
\end{equation}
Analogously, we zoom into these horizons by performing the change of variables 
\begin{equation}
    \varphi=\frac{A\tilde \varphi}{\rho\mathcal{K}^{3/2}\delta }, \qquad \theta=\theta_0+\delta \cosh \Xi,
    \label{angular zoom in super nariai}
\end{equation}
where $\mathcal{K}$ is defined by \eqref{K Nariai}.

Then, by taking the $\delta \rightarrow0$ limit, we obtain
\begin{equation}
    ds^2=\frac{r_A^2}{(r_A-r\rho)^2}\left(-f(r)dt^2+\frac{dr ^2}{f(r)} + \left(\frac{r}{\rho \sqrt{\mathcal{K}}}\right)^2\left(d\Xi ^2 +\sinh^2\Xi d\tilde \varphi^2 \right)\right).
\end{equation}
Thus we surprisingly find that the topology of a constant $t$ and $r$ slice is in fact conformal to a 2-hyperboloid of radius $r/(\rho\sqrt{\mathcal{K}})$. This result is quite interesting as we have just found a black hole characterized by hyperbolical symmetry instead of the more generally used spherical symmetry.

Additionally, we could perform both transformations simultaneously. However, this would mean reaching conformal infinity where the metric would blow up. Instead, we can analyze the geometry of the conformally rescaled metric
\begin{equation}
    \tilde g_{\mu \nu}=\Omega^2(r,\theta)g_{\mu \nu},
\end{equation}
where performing \eqref{angular zoom in super nariai} and \eqref{radial time zoom in super nariai} simultaneously gives 
\begin{equation}
    d\tilde s^2=\frac{1}{\mathcal{K}}\left(-\sin^2 \upsilon d\tau^2+d\upsilon ^2 + d\Xi ^2 +\sinh^2\Xi d\tilde \varphi^2 \right).
    \label{Super accelerating Nariai metric both horizons}
\end{equation}
And we conclude that the metric \eqref{Super accelerating Nariai metric both horizons} induces the topology of $dS^2\times H^2$\footnote{$H^2$ represents the 2 dimensional hyperboloid surface}. As for every point on the 2-hyperboloid, there is a two dimensional de Sitter spacetime.

\FIGURE{
    \centering
    \includegraphics[width=0.8\linewidth]{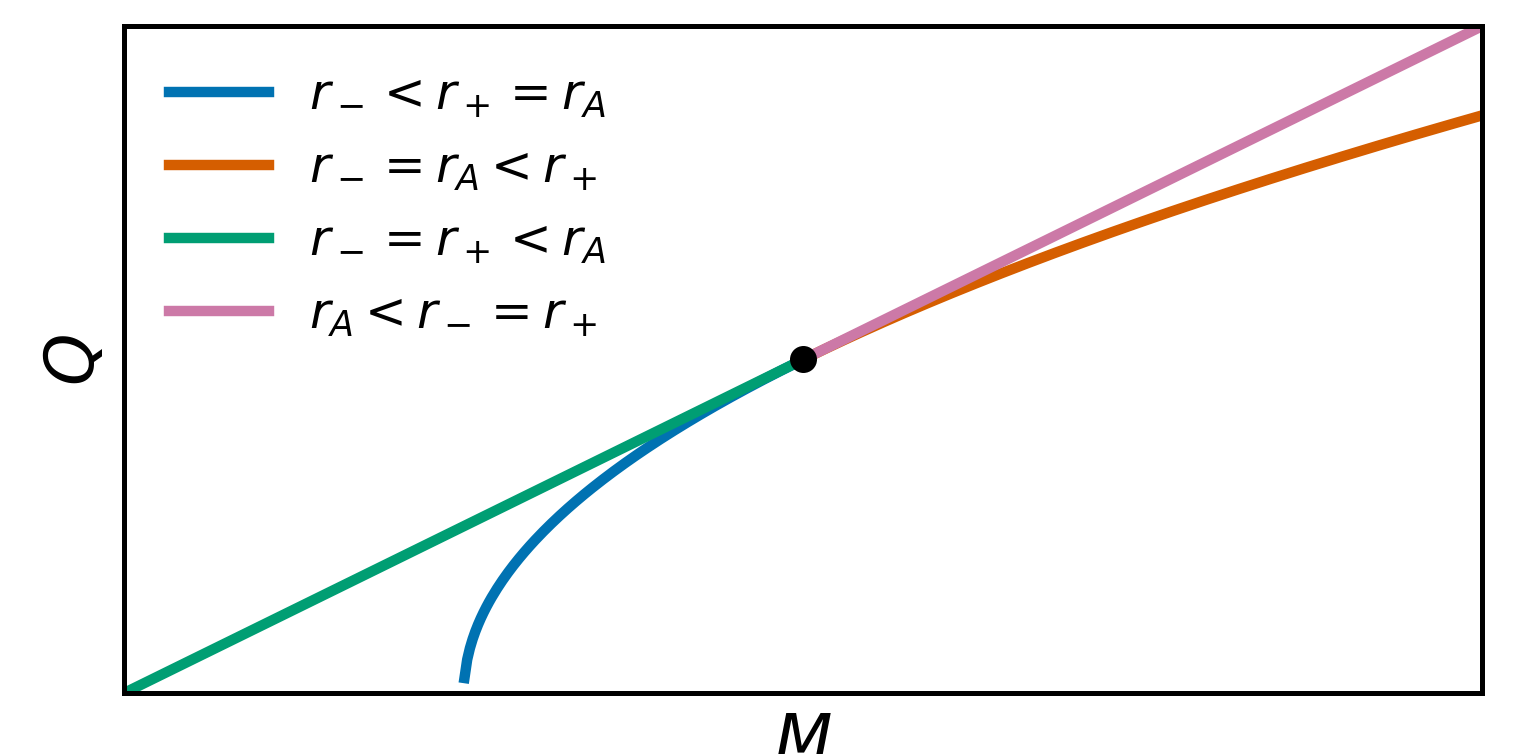}
    \caption{Charge and mass configuration for the different extremal configurations. The blue line shows the accelerating Nariai limit, the brown/orange line shows the super accelerating Bertotti Robinson limit, the green line shows the accelerating Bertotti Robinson and the magenta line shows the super accelerating Nariai Limit. At the point $M=Q=1/A$ all configurations merge and form the triple horizon limit.}
    \label{Phase space of extremal configurations}}
\section{Conclusions and open questions}\label{Conclusion}
\noindent
In this article we have explored the super accelerating regimes of the C-metric. These limits are obtained when the acceleration of the black hole is sufficiently high, such that the radius of the horizon associated with the acceleration of the black hole $r_A$ becomes smaller than the previous outer horizon of the black hole $r_+$. When charge is present on the spacetime we can obtain a second super accelerating limit where the radius $r_A$ is now less than the previous inner horizon $r_-$. 

These regimes were mostly excluded from the literature since they introduce a region $\theta$ in which the signature of the metric changes. This signature change introduces pathologies on the spacetime such as closed timelike curves that have been described in \ref{First super accelerating regime} as well as permits a configuration where the metric changes completely by an overall sign. We have addressed this problem by restricting the usual range of the angular coordinate, \((0,\pi)\), to a smaller interval and thereby introducing a new axis of symmetry achieved by a constant coordinate $\theta$, where the azimuthal Killing vector becomes null. We have shown that null geodesics cannot cross the resulting boundary in a finite affine parameter. This is related to the fact that, on a constant time and constant \(r\) slice of the manifold in the super accelerating regime, the region beyond this horizon forms a closed manifold that is topologically equivalent to a deformed 2-sphere, except at a single unavoidable point associated with a string or strut. Moreover, we inferred that when approaching the constant $r$ horizon surface where conformal infinity is reached, the manifold has the topology of $\mathbb{R}^3$. This corresponds to the cosmological horizon surface in both the regular and super accelerating regimes, except in the second super accelerating regime with $\theta\in(0,\theta_-)$, where conformal infinity is instead reached at the black hole event horizon. 

The emergence of these new axes of symmetry enriches the structure of the C-metric. While the causally accessible region contains only two conical singularities, one of which can be removed by an appropriate choice of the period of the azimuthal coordinate \(\varphi\), the full spacetime contains three conical singularities in the first super accelerating regime and four in the second. In either case, only one of these singularities can be removed by adjusting the period of \(\varphi\). Yet, it is still unclear whether the conical singularities outside of the causal region can also be thought of as providing the acceleration to the black hole.
Additionally, we have found that if the spacetime has a non zero cosmological constant, it is possible to conceive configurations with a naked singularity accompanied by or even without a cosmological horizon.

The computation of the mass of C-metric has been a subject of some problems mainly due to spacetime not being flat at high distances from the hole. For instance, it is not clear where the normalization of the time translation Killing vector should be performed and consequently, one finds that the Komar mass might be missing a multiplicative factor. On the regular accelerating regime however, it has been argued that it should be given by \eqref{Komar mass common accelerating regime} \cite{Griffiths:2006tk}. Using the same normalization, we have computed the Komar mass for the super accelerating regimes. We find that on the region $\theta\in(\theta_+,\pi)$, the Komar mass is independent of any constant $r$ surface the integral is performed at, and provided that one removes the conical singularity at the south pole, it is simply given by $(4A)^{-1}$. This is quite an intriguing result as it shows that the Komar mass of the black hole is dependent only on its acceleration while the mass parameter $M$ is absent in the contribution. In addition, if one considers the region $\theta\in(0,\theta_-)$ on the second super accelerating regime, and provided that the conical singularity at $\theta=0$ is removed, one finds that the Komar mass is negative and simply given by $-(4A)^{-1}$. The actual reason for this is unknown, but it might be an indication that the region $\theta \in (0,\theta_-)$ is pathological. Moreover, we have computed the area of the event and cosmological horizon where we have found that due to conformal infinity, the area of the cosmological horizon diverges for the regular and super accelerating regimes with $\theta \in (\theta_+,\pi)$. On the second super accelerating regime and $\theta\in (0,\theta_-)$, we find that the area of the cosmological horizon is finite while it is the area of the event horizon that diverges.

As a natural consequence of the super accelerating regime emerges new super accelerating extremal limits. These generate new solutions of the Einstein Maxwell field equations and have been obtained by applying a coordinate transformation that zooms into the extremal horizons. We have named them as the super accelerating Bertotti-Robinson and super accelerating Nariai spacetimes. Therefore, we have obtained the metric solution and electromagnetic field strength tensor that defines them as well as studied the topology they perform. Additionally, we found that on the super accelerating Nariai limit there is a configuration where the new axes of symmetry defined by $\theta=\theta_+$ and $\theta=\theta_-$ meet. There, by performing an appropriate coordinate transformation that zooms into this region, we found a black hole solution which presents hyperbolic symmetric instead of the more usual spherical symmetry.

The super accelerating regimes raise interesting open questions that are worth exploring, one of which concerns the quasinormal modes of the spacetime. These modes provide important information about its stability and spectroscopic properties. The quasinormal modes of the C-metric have been computed both numerically and analytically \cite{Destounis:2022rpk, Lei:2023mqx, Destounis:2020pjk, BarraganAmado:2023wxt, Chen:2024rov, Moura:2026ken}. Most analytical methods, such as the WKB method employed in \cite{Moura:2026ken}, can also be applied to the super accelerating regime, and some of the corresponding quasinormal modes have therefore already been calculated. However, we can now identify new families of quasinormal modes of the C-metric that have not yet been considered in \cite{Destounis:2020pjk}. Not only a new photon surface family becomes present on the super accelerating regime, but there are additional new near extremal families as well.
Moreover, quasinormal modes also play an important role on the stability of the Cauchy horizon, which are generically conjectured to be unstable. Even so, possible counter examples have already been found for spacetimes with a positive cosmological constant \cite{Cardoso:2017soq, Dias:2018etb} and for the C-metric as well \cite{Destounis:2020yav, Zhang:2023yco}. 
Since the super accelerating regime has the job of swapping the role of the horizons $r_+$ ,$r_-$ and $r_A$, it is curious to see whether the results found in \cite{Destounis:2020yav, Zhang:2023yco} generalize to the super accelerating regime. 
These kind of questions will be addressed in an upcoming article which is under preparation.


Additional open questions regarding the super accelerating regime include, among other aspects, questions related to black hole thermodynamics. Indeed, one might argue that the super accelerating regime cannot be reached from the regular accelerating one due to the third law of black hole mechanics. However, counter examples to this law have recently been found \cite{Crump:2026kgu, Kehle:2024vyt}. Actually, there might be a more convincing argument which prevents a continuous process shifting from one accelerating regime to the other. It is the fact that, to go, for instance and without loss of generality, continuously from the regular accelerating regime to the first super accelerating regime, there must be a configuration in which the cosmological horizon meets the event horizon. However, the topology of these two surfaces is different: whereas one is compact, the other is not. Thus, at the extremal (Nariai) configuration, the topology of the constant $t$ and constant $r$ surfaces must differ from that of the corresponding surfaces on either side of the transition. This implies a change in the topology of the spacetime as one passes continuously between the two regimes, raising the question of how such a topology change can be consistently described within general relativity \cite{Visser:1995cc}.

Furthermore, obtaining the laws of black hole thermodynamics of the C-metric has been a difficult problem to tackle. Many works have approached it either by considering a negative cosmological constant and the slow acceleration limit where the cosmological horizon is destroyed by the negative cosmological constant \cite{Hale:2025veb, Appels:2016uha, Abbasvandi:2018vsh}, essentially giving us only one horizon for the black hole which would not be in thermal equilibrium unless it was isolated. On the other hand, there is the work of \cite{Luo:2024cwm} where the laws of black hole thermodynamics of the C-metric in the Nariai limit, where the two horizons are in close contact and thus in thermal equilibrium were obtained. However, one has to deal with the problem of non compactness of the acceleration horizon. Ideally, but maybe not easily realizable, it would be interesting to investigate the thermodynamics of the super accelerating regime, where possibly our discovery that the Komar mass is only dependent on the acceleration of the black hole might have some relevance. Additionally, it is worth emphasizing that a slow super-accelerating regime is possible provided that the inequalities \(r_+>r_A>L\) are satisfied. Within this regime, a subset of configurations exists in which no cosmological horizon is present.

\paragraph{Acknowledgments:}
\noindent
I am grateful to my supervisor, Filipe Moura, for his continuous guidance and for introducing me to the field of black hole physics, and more specifically to the C-metric. I would also like to thank Adolfo Cisterna for reading an early draft and providing valuable feedback. I would also like to thank Professor José Pedro Mimoso for taking the time to discuss the C-metric with me, and Professor Francisco Lobo for his interest in and continued engagement with this work. Finally, during the EREP 2026 conference, I had the opportunity to discuss this work with several researchers. In particular, I would like to thank Adolfo Cisterna, Robie Hennigar, and Maxime Gardioux for engaging in fruitful discussions and providing valuable comments.

\end{document}